# Structure and magnetic properties of Heusler alloy $Fe_2NiZ$ (Z=Al, Ga, Si and Ge)

Y. J. Zhang (张玉洁), W. H. Wang (王文洪), H. G. Zhang (张红国), E. K. Liu (刘恩克), R. S. Ma (马瑞松) and G. H. Wu (吴光恒)[*]

*Beijing National Laboratory for Condensed Matter Physics, Institute of Physics, Chinese Academy of Sciences, Beijing 100190, People's Republic of China*

**Abstract**

The Heusler alloys $Fe_2NiZ$ (Z=Al, Ga, Si and Ge) have been synthesized and investigated focusing on the phase stability and the magnetic properties. The experimental and theoretical results reveal the covalent bonding originated from *p-d* hybridization takes an important role in these alloys, which dominates the stability of ordered structure but leads to the decline of the band splitting. The electronic structure shows the IV group main group element (Si and Ge) provides stronger covalent effect than that of the III group element (Al and Ga). It has been found that the variations of the physical parameters, lattice constants, critical ordering temperature, magnetic moments and Curie temperature, precisely follow these covalent characters.

## 1. Introduction

Heusler alloys have attracted widely attention for its various applications as well as their remarkable physical properties [1]. Heusler alloy is also a typical object for studying the ordered atomic configuration and magnetic coupling for intermetallic compounds[2]. Therefore, synthesizing and investigating new systems becomes an attractive subject for investigators working in this field. Among the $Fe_2$-based Heusler alloys, the $Fe_2CrZ$ (Z is the main group elements)[3, 4] and $Fe_2MnZ$[5, 6] alloys exhibit very interesting magnetic structure and magnetic martensitic transformation shape memory properties. For $Fe_2NiZ$ systems, there are also some reports related to the phase stability and the magnetic properties[3, 7, 8]. However, the detail about the relationship between the phase stability and the magnetic properties for this system has not been investigated sufficiently.

---

[*] Correspond author. Tel.: +86 010 82649247; Electronic mail: ghwu@iphy.ac.cn.

In this paper, the structure and magnetic properties are systematically investigated on Heusler alloys $Fe_2NiZ$ (Z=Al, Ga, Si and Ge). We found that, selecting Al, Ga, Si and Ge as Z, four kinds of Heusler alloys with well ordered $Hg_2CuTi$-type structure can be synthesized. A series of physical parameters, such as lattice constants, magnetic moments, Curie temperature and ordering temperature have been measured from those samples. In order to study the relation between the magnetic properties and the chemical bonding in this $Fe_2NiZ$ system, the theoretical calculation for the band structure and electron localization function has been performed. Our results indicate that the covalent bonding established by the *p-d* orbital hybridization between the atoms of main group and 3*d*-elements takes an important role for the various physical properties in the $Fe_2NiZ$ alloys.

2. Experimental and computational details

We synthesized the $Fe_2NiZ$, Z = Al, Ga, Si and Ge compounds and studied the structure,

magnetic properties, phase stability and electronic structures by experiment and calculating methods. Stoichiometrical $Fe_2NiZ$ were prepared by arc-melting the elements with the purity higher than 99.9% and subsequently, thermal treatment (1000 $^oC$ and 72 hrs) was used to homogenize the samples. In order to ensure high chemical ordering, the homogenized ingots were further annealed at 650 $^oC$ for 72 hrs, and subsequently quenched in an ice-water mixture. The lattice structure was identified by X-ray diffraction (XRD) technique with Cu-Kα radiation (λ=1.5418 Å). The differential scanning calorimeter (DSC) method was used to investigate the thermal stability of the samples, with the cooling/heating rate of 10 $^oC$/min. The magnetic measurements were performed using the SQUID magnetometer (Quantum Design). High temperature AC magnetic susceptibility was measured to determine the Curie temperature, $T_C$. Calculations about the energy band structures and magnetic moments were performed by the KKR-CPA-LDA method[9-12]. The electron distribution was obtained through the study of the electron localization function (ELF)[13-15].

## 3. Results and discussions

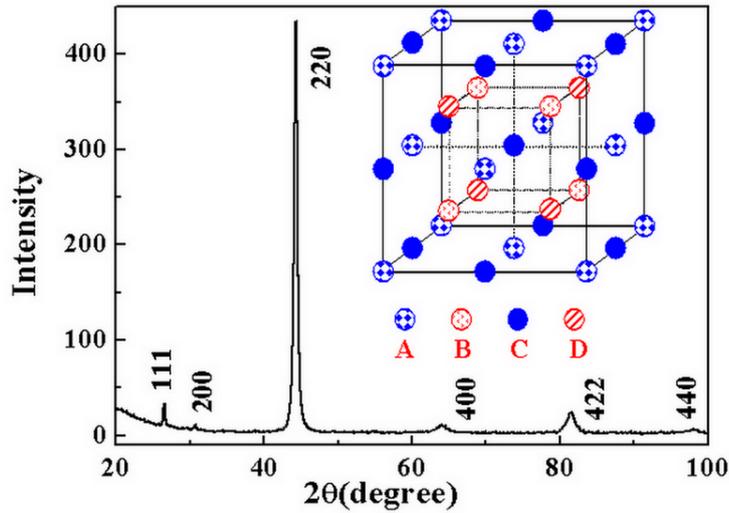

Fig. 1. (color online) The XRD pattern of $Fe_2NiAl$. The inset is visualized diagram of the structure of Heusler alloys with the signed equivalent lattice sites.

The XRD pattern of $Fe_2NiAl$ has been shown in Fig. 1, which fits well with the body centered cubic (bcc) structure. The other $Fe_2NiZ$ alloys have the similar XRD patterns. The superlattice reflection peaks (111) and (200) can be observed, that means, the $Fe_2NiZ$ samples have a highly chemical ordered structure. The inset of Fig. 1 is a schematic of Heusler lattice with four equivalent lattice sites $A$ (0, 0, 0), $B$ (0.25, 0.25, 0.25), $C$ (0.5, 0.5, 0.5) and $D$ (0.75, 0.75, 0.75). It has been indicated that the covalent bonding, which is mainly originated from $p$-$d$ hybridization[16, 17] between the main group element and the transition metal, has great influence on the properties and the atomic arrangement of alloys[18]. Thus, an empirical rule relying on the valence electrons[19, 20] indicates that the atoms with relatively more valence electrons tend to occupy the $A$ and $C$ sites preferentially, while those with relatively fewer valence electrons will occupy the $B$ and $D$ sites. It is consistent with this empirical rule for $Fe_2NiZ$ alloys structuring in a $Hg_2CuTi$ structure, in which $A$ and $B$ sites are occupied by Fe atoms, $C$ and $D$ sites by the atoms of Ni and main group elements[21, 22]. Experimentally, the intensity of (111) peak is higher than that of (200) in Fe$_2$NiAl, corresponding to the character of Hg$_2$CuTi ordered structure[23]. The alloys of Fe$_2$NiSn and Fe$_2$NiSb were also synthesized and a bcc main phase could be identified from the XRD patterns, but quite large amount of second phase can also been observed. The second phase might be eliminated by some special heat treatments. These results imply that Fe$_2$NiZ may be a big family of Heusler alloys.

The lattice constants of Fe$_2$NiZ (Z = Al, Ga, Si and Ge) samples have been obtained by refining the XRD patterns, as listed in Table 1. One may see that the lattice constants show little variation when Z = Al, Ga, Ge whereas the Fe$_2$NiSi is much smaller than the others. Considering the covalent radii of the main group atoms (Al, Ga and Ge are of around 1.21 Å and Si is of 1.11 Å), the lattice constants may reflect the covalent effect which dominates in the structure of these Heusler alloys.

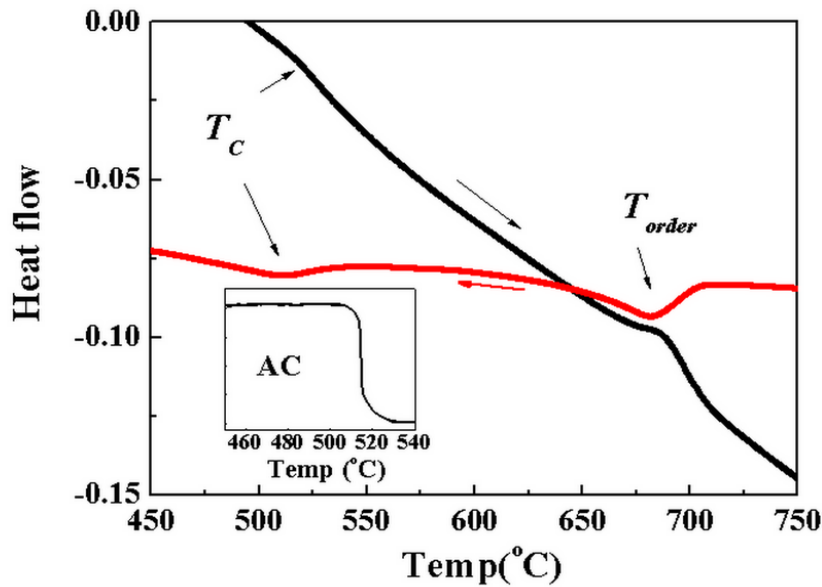

Fig. 2. (color online) The heating and cooling curves of Fe$_2$NiGa measured by DSC. The rate of heating and cooling is 10°C/min. The inset is the AC susceptibility curve showing $T_C$.

Fig. 2 presents the DSC curves for Fe$_2$NiGa alloy. Two sets of endothermic and exothermic peaks can be clearly observed at the temperature around 513 °C and 686 °C. The inset of Fig. 2 shows a spontaneous magnetization behavior observed by AC susceptibility measurement. Therefore, the heat events occurred at 513 °C can be attributed to the spontaneous magnetization, showing the Curie Temperature ($T_C$) of $Fe_2NiGa$. The sharp peaks around 686 °C shows a critical temperature ($T_{order}$) for atomic ordering, through which the $Fe_2NiGa$ system transforms between the $Hg_2CuTi$ and B2 structures, agreeing with the previous report in Ref. [7]. The $T_C$ and $T_{order}$ of four alloys are listed in Table 1. It can be found that the $T_{order}$ varies depending strongly on the species of Z element. $Fe_2NiGa$ has the lowest $T_{order}$ of 686 °C, while $Fe_2NiSi$ has the highest one of 822 °C. On the other hand, the $T_C$ of $Fe_2NiZ$ shown in Table 1 presents a distinct behavior: $Fe_2NiGa$ has the much higher $T_C$ than that of $Fe_2NiSi$.

Generally, the $T_{order}$ reflects the phase stability of the system. As we mentioned above, the *p-d* hybridization between the main group elements and the transition metals arranges the atoms occupying the non-equivalent sites[18, 24]. In light of this consideration, the covalent bonding effect from III group elements ($Z_{III}$) should be weaker than that of IV group elements ($Z_{IV}$), and the strength of covalent bonding tends to decrease in the sequence Si> Ge> Al> Ga approximately. Additionally, it was believed that the magnetic moments and internal fields in Heusler alloys can be weakened by the covalent bonding[25]. This is the reason why the $T_C$ has the trend distinct with the $T_{order}$.

The molecular and atomic moments obtained by experimental and calculating methods are presented in Table 2. The molecular moments of 4.10~4.89 $\mu_B$ for four alloys have been measured experimentally. The calculated molecular moments match well with the experimental ones. It can be seen that these results do not agree with the Slater-Pauling rule of M = Z-24, as observed in $Co_2Fe$-based Heusler alloys where the molecular moments of $Co_2FeAl$ and $Co_2FeSi$ are of about 5

$\mu_B$ and 6 $\mu_B$, respectively[26, 27]. In the present case, the moments of the alloys with $Z_{IV}$ are smaller than those of the alloys with $Z_{III}$.

Table 1. The lattice constant (a), critical ordering temperature ($T_{order}$) and Curie temperature ($T_C$) of Fe$_2$NiZ.

| Fe$_2$NiZ | a (Å) | $T_{order}$ (°C) | $T_C$ (°C) |
| --- | --- | --- | --- |
| Al | 5.778 | 739 | 691 |
| Ga | 5.776 | 686 | 513 |
| Si | 5.671 | 822 | 484 |
| Ge | 5.761 | 798 | 412 |

Table 2. The molecular and atomic moments (in $\mu_B$) of Fe2NiZ alloys obtained by experimental measurements ($M_{exp}$) and calculations ($M_{cal}$).

| Z | $M_{exp}$ | $M_{cal}$ | Fe$A$ | Fe$B$ | Ni$C$ | Z$D$ |
| --- | --- | --- | --- | --- | --- | --- |
| Al | 4.46 | 4.71 | 1.74 | 2.67 | 0.45 | -0.08 |
| Ga | 4.89 | 4.82 | 1.83 | 2.67 | 0.45 | -0.07 |
| Si | 4.10 | 4.46 | 1.61 | 2.58 | 0.34 | -0.06 |
| Ge | 4.38 | 4.72 | 1.77 | 2.65 | 0.37 | -0.04 |

Table 2 shows Fe$A$ and Fe$B$ are the main contributors for the ferromagnetic structure of Fe$_2$NiZ. The atomic moment of Fe atoms is of 1.61-1.83 $\mu_B$ in $A$ site whereas 2.58-2.67 $\mu_B$ in $B$ site in the present four alloys. The smaller moment of Fe in $A$ site can be attributed to the reduction of intrinsic magnetic moment by hybridizing with Z atoms[25]. The ordering dependence of molecular moment has also observed experimentally in the present work. Without annealing below the $T_{order}$, the homogenize Fe$_2$NiGa sample has a low moment of 4.16 $\mu_B$, indicating a random occupation of the Fe atoms.

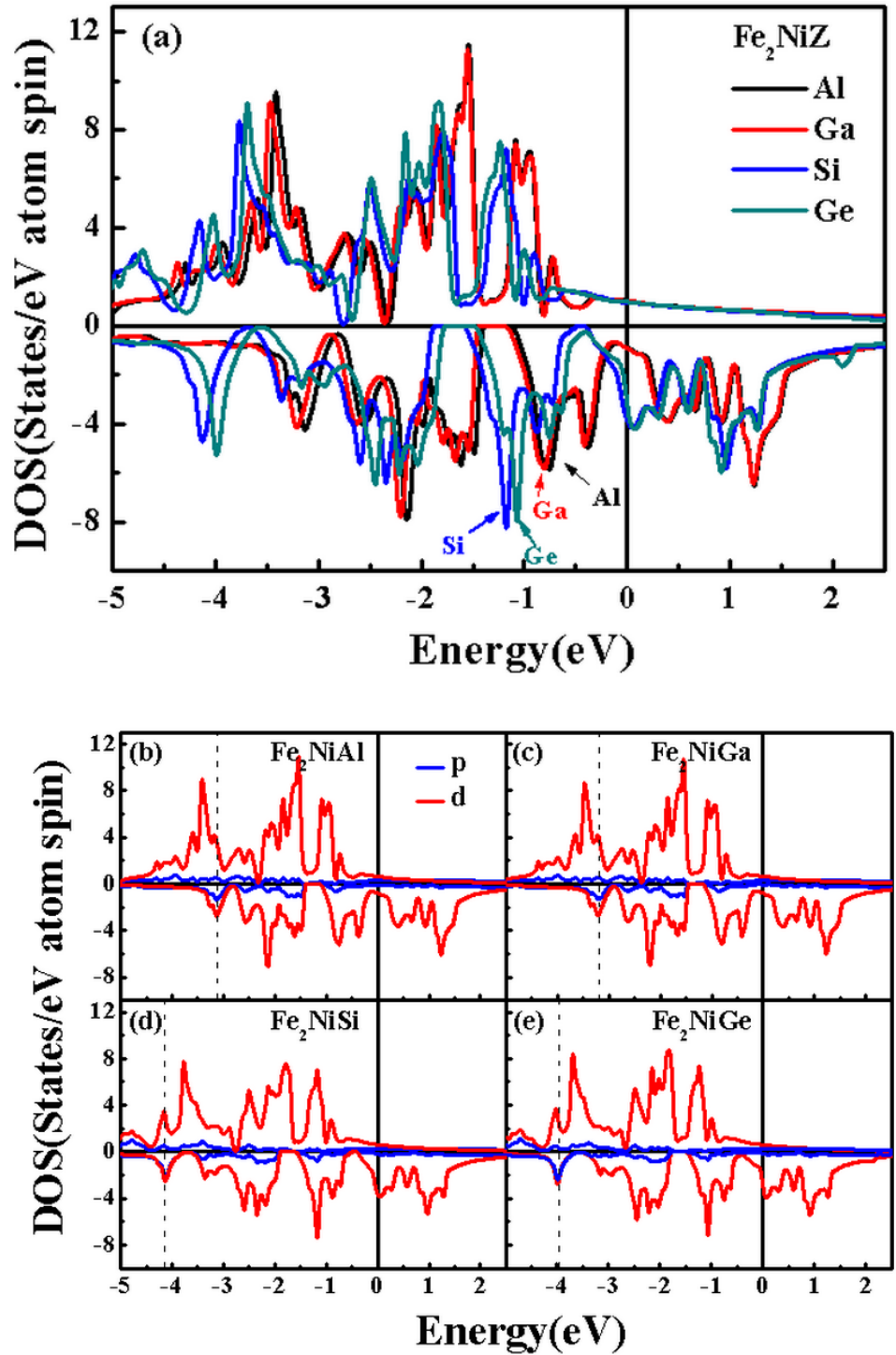

Fig. 3. (color online) The total DOS of Fe$_2$NiZ (Z = Al, Ga, Si and Ge) (a) and the partial DOS of *p*- and *d*- electrons (b)-(e).

Our calculated electronic structure indicates that, in these systems, the $p$ electrons of Z element (Al, Ga, Si and Ge) hybridize energetically with the $d$ electrons of the transition metals. Fig. 3 plots the electronic density of states (total DOS) of $Fe_2NiZ$ (Z =Al, Ga, Si and Ge) (a) and the partial DOS of $p$ and $d$ electrons (b)-(e). The bands are formed mostly of $d$-states on Fe and Ni, while $p$-states of Z elements participate in the hybridization[28]. The feature of distinct valleys in the DOS indicates the existence of the covalent interaction[29, 30]. It looks that the alloys with the same main group elements have the similar total DOS. As indicated in Fig. 3 (a), the peaks shift to the lower energies when the Z elements become the $Z_{IV}$, which is more obvious for spin-down states. This implies a decline on band splitting.

As shown in Fig. 3 (b)-(e), there are visible $p$-$d$ hybrid peaks of the spin-down states below the $E_F$ around -3.2 eV (for Z = $Z_{III}$) or -4.2 eV (for Z = $Z_{IV}$), which are marked with dotted lines. One can see the $p$-DOS of the $Z_{IV}$ apparently shift to lower energy, showing stronger hybrid effect than those of $Z_{III}$. So the weakness of the magnetic coupling observed experimentally should be attributed to the $p$-$d$ hybridization[30]. These calculations reveal the competition between covalent interaction and magnetic coupling, which is the physical mechanism for the molecular moments of these alloys disobeying the Slater-Pauling rule.

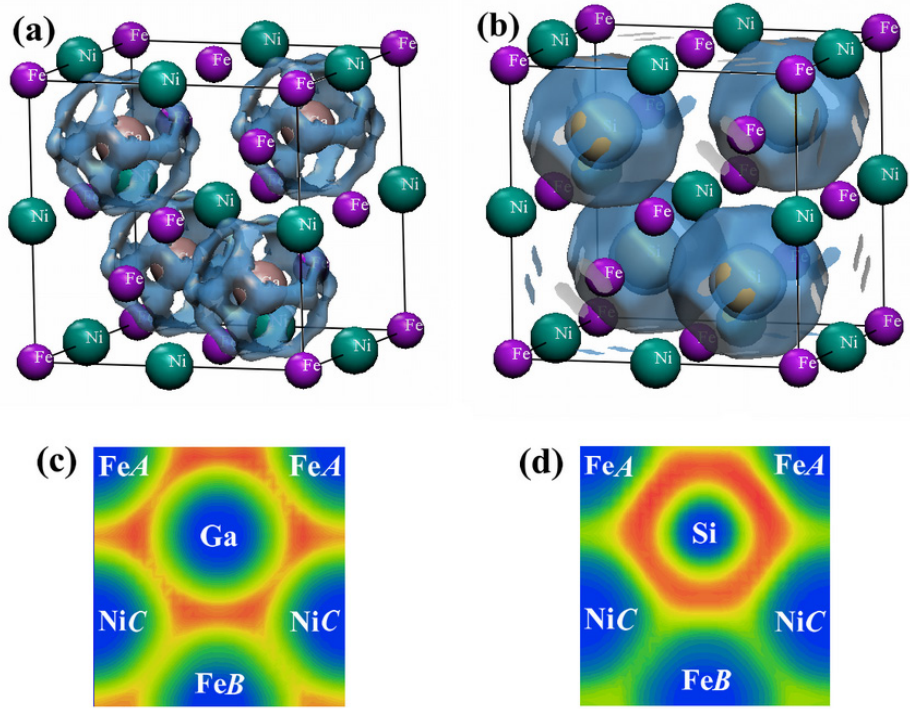

Fig. 4. (color online) The 3D ELF diagrams of $Fe_2NiGa$ (a) and $Fe_2NiSi$ (b), the constant ELF value is 0.006 and the 2D ELF of the alloys with $Z_{III}$ (c) and $Z_{IV}$ (d) on the (110) crystal plane; the linear color scale extends from 0 (blue) to 0.012 a.u. (red).

The electron localization function (ELF) has been usually used as an efficient method to identify the nature of chemical bonding in intermetallic compounds visually[13, 14, 31, 32]. Fig. 4 shows the three-dimensional (3D) and two-dimensional (2D) visualizations of ELF for $Fe_2NiGa$ and $Fe_2NiSi$, which can exhibit the different electron distribution between the alloys with $Z_{III}$ and $Z_{IV}$. In the 3D diagrams, the iso-surface corresponding to the constant ELF value of 0.006 represents for regions with specific space configuration. These regions distribute around the main group elements Ga and Si atoms can be taken as paired electron[14], including covalent bonds or lone pairs. The ELF value is relatively low but may play an active role for affecting the atomic configuration in the intermetallic compounds. Usually, the covalent bonds should locate on the line between Z-Fe$A$ and Z-Ni$C$ while lone pairs appear along the Z-Fe$B$ (next nearest) vector[33].

The 3D ELF shapes of the alloys with Ga and Si reflect the different strength of covalent bonds.

The high and low ELF values in 2D ELF graphs correspond to areas of localized electrons and spatial area around these maxima, respectively[33]. The regions showing the maximum of ELF in Fig. (c) and (d) is located around the main group element and along the bond axis of Z-Fe$A$ and Z-Ni$C$, confirming their significant shared character and the occurring of covalent bond. It illustrates that the covalent effect of Si is stronger than Ga. This matches the sequence of covalent bonds strength mentioned above. There is noticeable difference in the spatial configuration of ELF: the basin of alloys with $Z_{III}$ show a hexagonal shape, while it is almost sphere as the case containing $Z_{IV}$ (Fig. 4 (c) and (d)). This phenomenon is more obviously in 3D ELF where the Fe$_2$NiGa shows a coop-type iso-surface, but the Fe$_2$NiSi is the closed diamond-type with the same ELF value (Fig. 4 (a) and (b)). This distinction means the electrons of the $Z_{IV}$ element have more isotropic space configuration. One may find that a series of physical parameters observed in the present work, lattice constants, $M_S$, $T_C$, $T_{order}$ and the moment variation of Fe$A$, Ni$C$, as shown in Table 1 and Table 2, precisely follow to the strength of covalent effect.

## 4. Conclusion

The structural and magnetic properties of Heusler alloy Fe$_2$NiZ (Z = Al, Ga, Si and Ge) have been systemically investigated by experimental and theoretical methods. All these samples have ordered atoms in a Hg$_2$CuTi structure with the lattice constants dominated by the covalent radii of the main group element Z. They have quite high $T_C$ and molecular moment of more than $4\mu_B$. The theoretical calculation indicates that the Fe$B$ atoms are the main magnetic contributors and have much larger moments than that of the Fe*A*. The covalent behavior and the sequence of its strength in $Fe_2NiZ$ alloys have been confirmed by calculation and experiments. It has been found that the variations of the physical parameters observed in the present work, lattice constants, $T_{order}$, molecular moments and atomic moments and $T_C$, precisely follow the change of covalent characters. This indicates the important role of covalent bonding for the structural and magnetic properties of $Fe_2NiZ$ Heusler alloys.

## Acknowledgements

This work is supported by the National Natural Science Foundation of China in Grant No. 51031004 and 11174352 and National Basic Research Program of China (973 Program, 2010CB833102).